\documentclass[aps,pre,twocolumn,superscriptaddress,showpacs,floatfix,a4paper]{revtex4}

\usepackage{amsmath}
\usepackage{amssymb}
\usepackage{epsfig}
\usepackage{multirow}
\usepackage{graphicx}

\newcommand {\dd}{{\mathrm d}}

\begin{document}

\title{First-order layering and critical wetting transitions in
  non-additive hard sphere mixtures}

\author{Paul\ Hopkins}
\affiliation{H.H.\ Wills Physics Laboratory, University of Bristol,
  Tyndall Avenue, Bristol BS8 1TL, UK}

\author{Matthias\ Schmidt}
\affiliation{H.H.\ Wills Physics Laboratory, University of Bristol,
  Tyndall Avenue, Bristol BS8 1TL, UK}
\affiliation{Theoretische Physik II, Physikalisches Institut,
  Universit{\"a}t Bayreuth, D-95440 Bayreuth, Germany}

\date{17 January 2011, submitted to Phys. Rev. E, Rapid Communication}

\begin{abstract}
Using fundamental-measure density functional theory we investigate
entropic wetting in an asymmetric binary mixture of hard spheres with
positive non-additivity. We consider a general planar hard wall, where
preferential adsorption is induced by a difference in closest approach
of the different species and the wall. Close to bulk fluid-fluid
coexistence the phase rich in the minority component adsorbs either
through a series of first-order layering transitions, where an
increasing number of liquid layers adsorbs sequentially, or via a
critical wetting transition, where a thick film grows continuously.
\end{abstract}


\maketitle

Studying the interfacial properties of liquid mixtures is of
significant fundamental and technological relevance
\cite{bonn2009wetting}. Bulk liquid-liquid phase separation, which can
arise at or close to room temperature, is usually associated with rich
phenomenology of interfacial behaviour at a substrate. Gaining a
systematic understanding of how the different types of intermolecular
and of substrate-molecule interactions induce phenomena such as
wetting, layering, and drying at substrates constitutes a major
theoretical challenge. Relevant for surface adsorption of liquids are
Coulombic and dispersion forces, but also solvent-mediated and
depletion interactions which occur in complex liquids. Arguably the
most important source for the emergence of structure in dense liquids
is the short-ranged repulsion between the constituent particles; this
may stem from the overlap of the outer electron shells in molecular
systems or from screened charges or steric stabilization in colloidal
dispersions.

Hard sphere fluids form invaluable reference models for investigating
the behaviour of liquids at substrates. Both the pure
\cite{hansen2006tsl,roth2010fundamental} and binary
\cite{roth2000binary} hard sphere fluids are relevant, the latter
playing an important role when adding e.g.\ electrostatic interactions
in order to study wetting of ionic liquids at a substrate
\cite{oleksyhansen}. The most general binary mixture is characterized
by independent hard core distances between all different pairs of
species, and is referred to the non-additive hard sphere (NAHS)
model. Here the cross species interaction distance can be smaller or
larger than the arithmetic mean of the like-species diameters. The
NAHS model gives a simplified representation of more realistic pair
potentials, i.e.\ charge renormalisation effects in ionic mixtures in
an explicit solvent induce non-additive effective interactions between
the ions \cite{kalcher}. It is also a reference model to which
attractive or repulsive tails can be added \cite{referenceNAHS}. The
Asakura-Oosawa-Vrij (AOV) model of colloids and non-adsorbing polymers
\cite{asakuraoosawavrij} is a special case where one of the diameters
(that of the polymers) vanishes.

It is surprising that the wetting behaviour of the general NAHS model
is largely unknown, given the fundamental status of the model. In this
Letter we address this problem and consider the NAHS fluid at a
general, non-additive hard wall. We find a rich phenomenology of
interfacial phase transition, including two distinct types of surface
transitions: one is layering, where the adsorption of one of the
phases occurs through a number of abrupt jumps, and the other is
critical wetting, where the thickness of the adsorbed film grows
continuously when varying the statepoint along the bulk fluid-fluid
binodal. Via changing the wall properties a crossover between these
transitions occurs.

\begin{figure}[t]
\centering
\includegraphics[width=8cm]{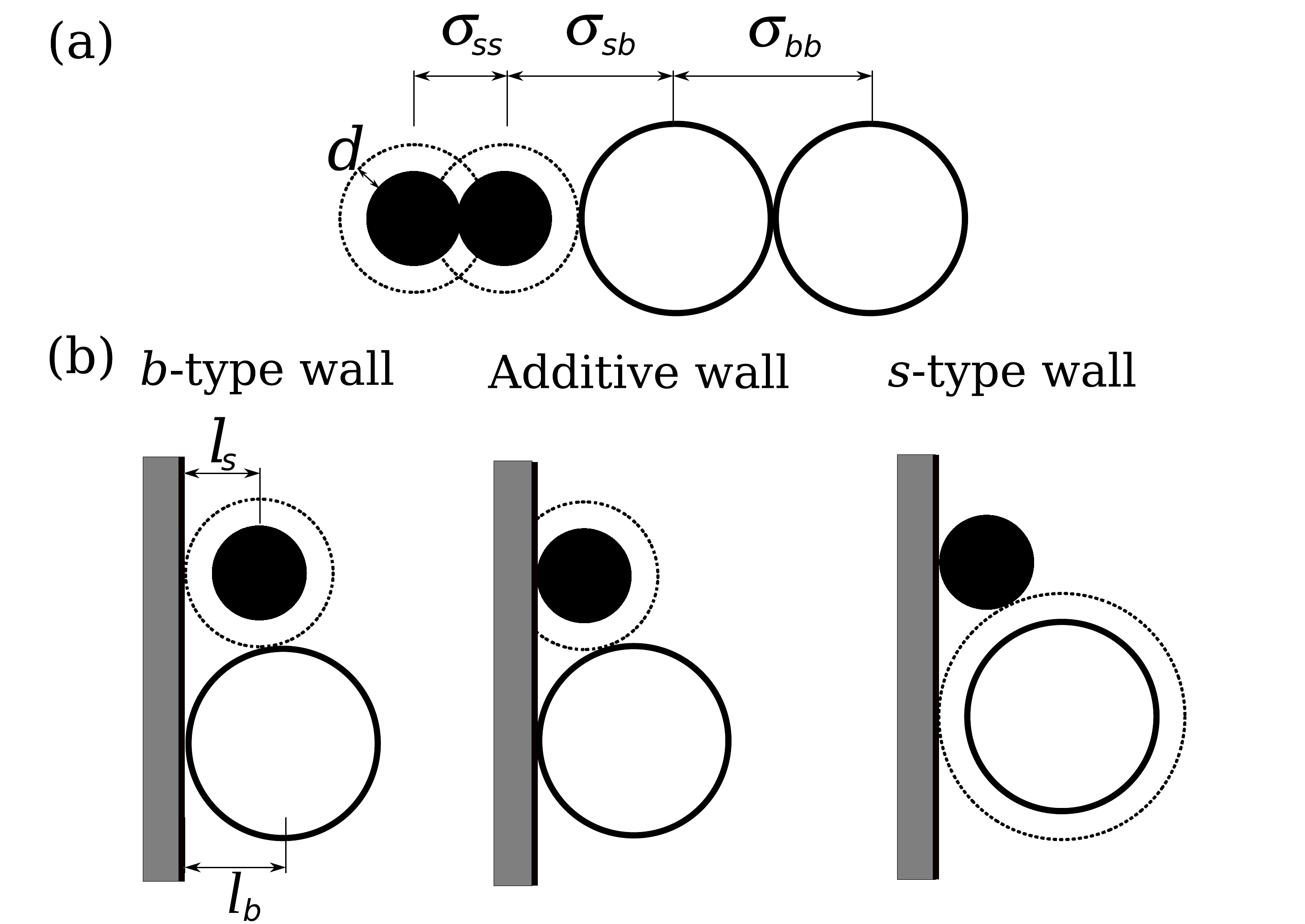}
\caption{\label{fig:diagram} (a) Illustration of the asymmetric NAHS
  model with positive non-additivity. The solid boundaries represent
  the hard cores of the small and big species. The dotted line
  represents the non-additive hard core between unlike species, which
  here is attributed only to smaller particles. (b) Three examples of
  general planar hard walls. The additive wall treats the two species
  equally, while the $b$-type and $s$-type walls have properties
  similar to the big and small particles, respectively.}
\end{figure}

The binary NAHS model is defined by the pair potentials
$v_{ij}(r)=\infty$ for $r<\sigma_{ij}$ and~$0$ otherwise, where $i,j$
= $s,b$ refers to the small and big species, respectively,
$\sigma_{ss}$ and $\sigma_{bb}$ are the diameters of the small and big
particles, respectively, and $r$ is the center-to-center distance. The
cross-species diameter is
$\sigma_{sb}=\tfrac{1}{2}(1+\Delta)(\sigma_{ss}+\sigma_{bb})$, where
$\Delta\geq-1$ measures the degree of non-additivity, see
Fig.~\ref{fig:diagram}(a) for an illustration of the length scales.
The model is characterised by the size ratio,
$q=\sigma_{ss}/\sigma_{bb}\leq1$, and by $\Delta$. In this Letter we
restrict ourselves to the asymmetric size ratio $q=0.5$, and to
positive non-additivity $\Delta=0.2$, as a representative case. We
relate $\Delta$ to a length scale via
$d=\tfrac{1}{2}(\sigma_{ss}+\sigma_{bb})\Delta\equiv\sigma_{sb}-
\tfrac{1}{2}(\sigma_{ss}+\sigma_{bb})$, where here
$d=0.3\sigma_{ss}$. The statepoint is characterised by two partial
bulk packing fractions, $\eta_i=\pi\sigma_{ii}^3\rho_i/6$, where
$\rho_i$ is the number density of species $i$.  We define a general
planar hard wall via the external potentials $u_i(z)=\infty$ if
$z<l_i$, and 0 otherwise; here $z$ is the distance between the wall
and the particle center, and $l_i$ is the minimal distance of approach
of species $i=s,b$. Clearly the origin in $z$ is irrelevant, so the
only further control parameter is the wall offset, $\delta
l=l_b-l_s$. For additive hard sphere mixtures it is common to set
$l_i=\sigma_{ii}/2$; for our model parameters this results in $\delta
l=0.5\sigma_{ss}$. Besides this `additive wall', two further special
cases are shown in Fig.~\ref{fig:diagram}(b). The $b$-type wall has
properties similar to the big particles so that it sees these with
their `intrinsic' size $l_b=\sigma_{bb}/2$, but sees the small
particles with their `non-additive' size $l_s=\sigma_{ss}/2+d$, such
that $\delta l=0.2\sigma_{ss}$. We expect that the bigger particles
adsorb more strongly to this wall.  Conversely, the $s$-type wall has
properties similar to the small particles, so that it sees these with
their `intrinsic' size $l_s=\sigma_{ss}/2$, and sees the big particles
with their `non-additive' size $l_b=\sigma_{bb}/2+d$, so that $\delta
l=0.8\sigma_{ss}$. Thus, one expects the small particles to adsorb
more strongly.

We investigate the inhomogeneous NAHS fluid using a fundamental
measure density functional
theory~\cite{schmidt2004rfn,hopkins2010binary}. Comparison of
theoretical results to Monte Carlo simulation data for bulk
fluid-fluid phase diagrams \cite{schmidt2004rfn,hopkins2010binary},
partial radial distribution functions
\cite{schmidt2004rfn,ayadim2010generalization} and density profiles in
planar slits \cite{hopkins2010all} indicates very good quantitative
agreement. We obtain equilibrium density distributions $\rho_i(z)$
from the grand potential functional, $\Omega[\rho_s,\rho_b]$, by
numerical solution of $\delta\Omega/\delta\rho_i(z)=0$, $i=s,b$. To
calculate coexisting (bulk or surface) states we use the equality of
the chemical potentials $\mu_s$, $\mu_b$, and $\Omega$ in the two
phases. The NAHS functional~\cite{schmidt2004rfn} features both a
large number of terms and a large number of convolutions that take
account of the non-locality. Therefore the accurate calculation of
density profiles close to phase coexistence, and close to interfacial
transitions, is a challenging task.

\begin{figure}[t]
\centering
\includegraphics[width=8.0cm]{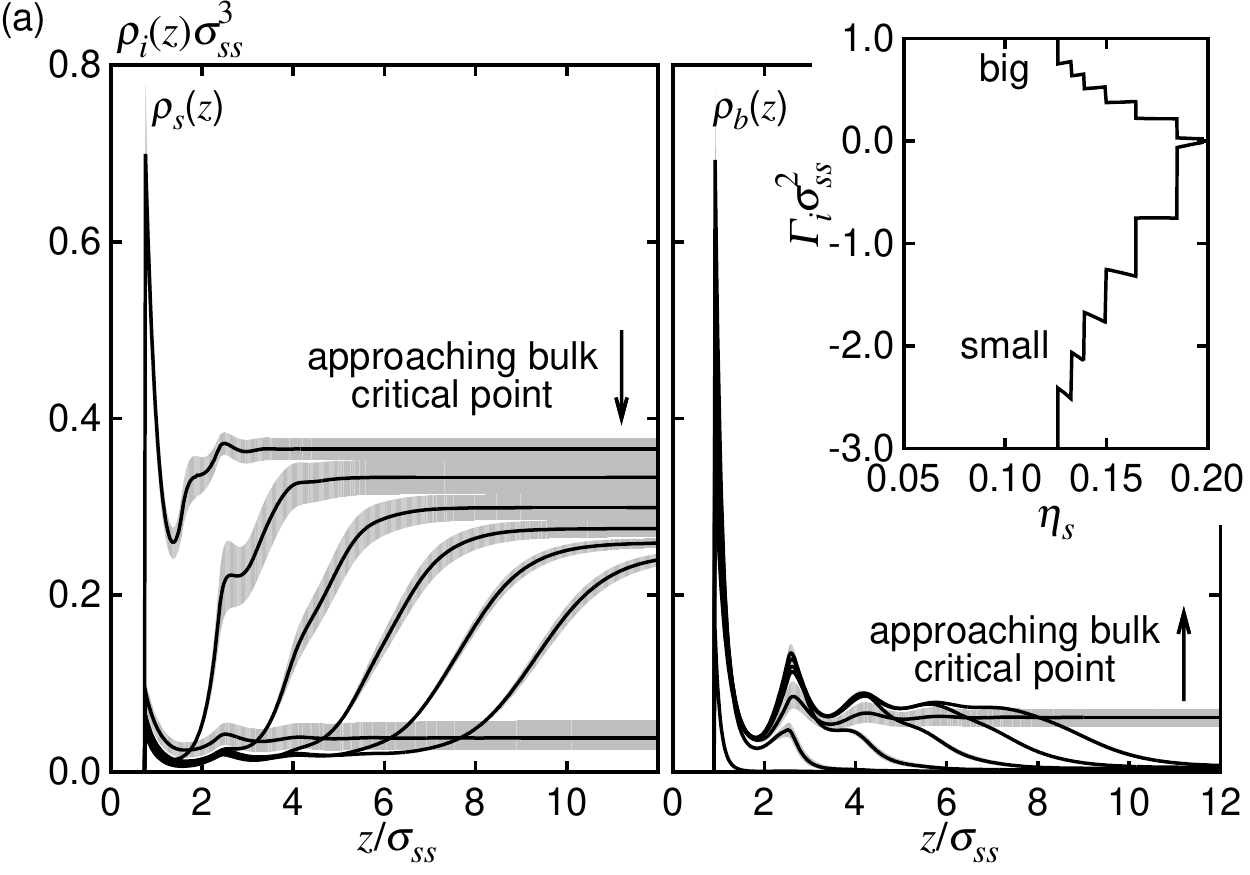}
\includegraphics[width=8.0cm]{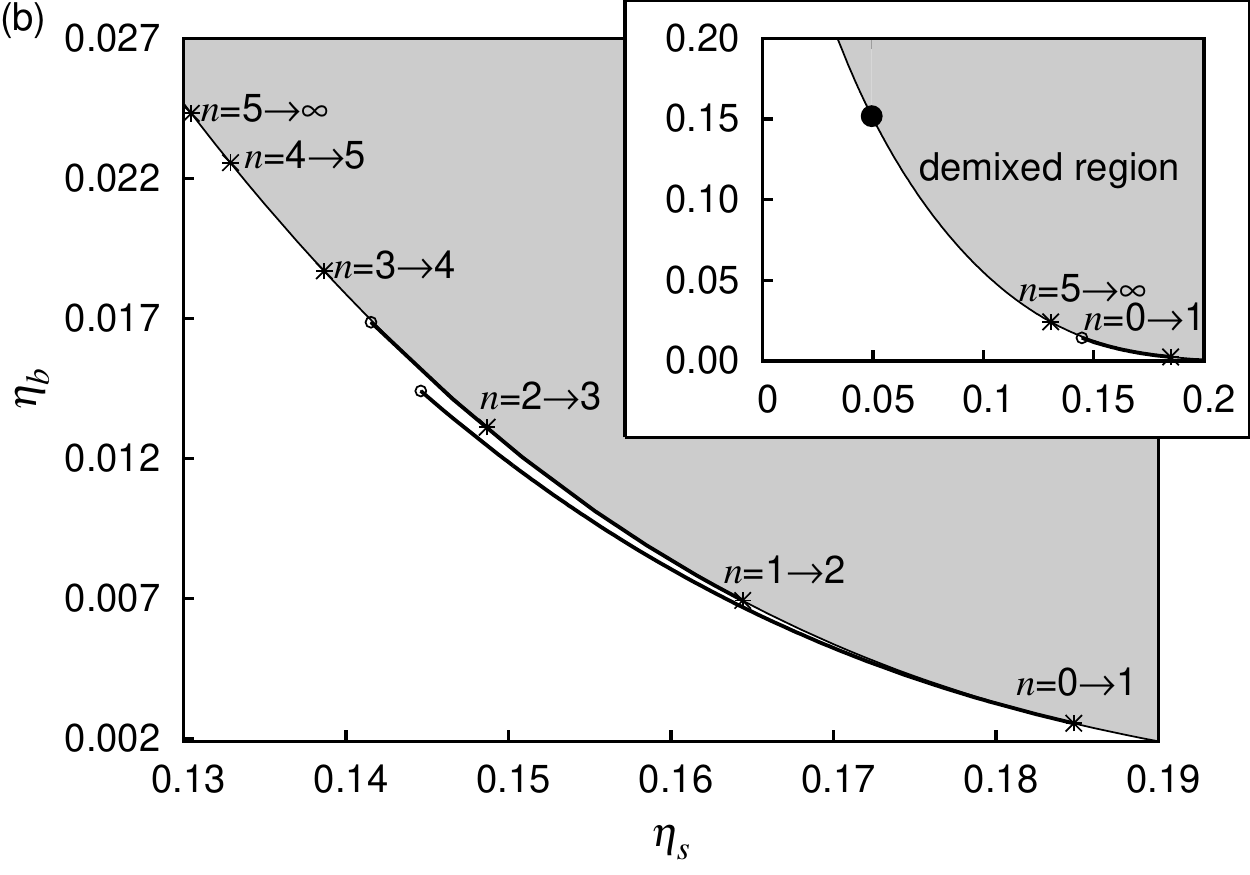}
\caption{\label{fig:1w_profiles} (a) Pairs of density profiles,
  $\rho_s(z)$ (left panel) and $\rho_b(z)$ (right panel), of the NAHS
  fluid with $q=0.5$ and $\Delta=0.2$ at a $b$-type wall and at bulk
  coexistence on the $s$-rich side of the phase diagram. The shaded
  regions represent the range of profiles possessing $n=0$ or
  $1,2,3,4$ or $5$ adsorbed $b$-rich layers, and the region where the
  adsorbed film becomes infinitely thick. The solid lines represent
  specific examples from the middle of each range. The inset shows the
  adsorption of each species, $\Gamma_i$, as a function of
  $\eta_s$. (b) The corresponding phase diagram in the
  ($\eta_s,\eta_b$) plane. There is a series of layering transitions
  that intersect the bulk binodal ($\ast$) and descend into the
  $s$-rich one phase region, ending in a surface critical point
  ($\circ$). For clarity only the first two transitions are shown in
  full, while the remaining transitions are represented only by their
  intersection with the bulk binodal. The inset shows the location of
  the first layering and the `wetting' transitions in relation to the
  bulk critical point ($\bullet$).}
\end{figure}

\begin{figure}[t]
\centering
\includegraphics[width=8.0cm]{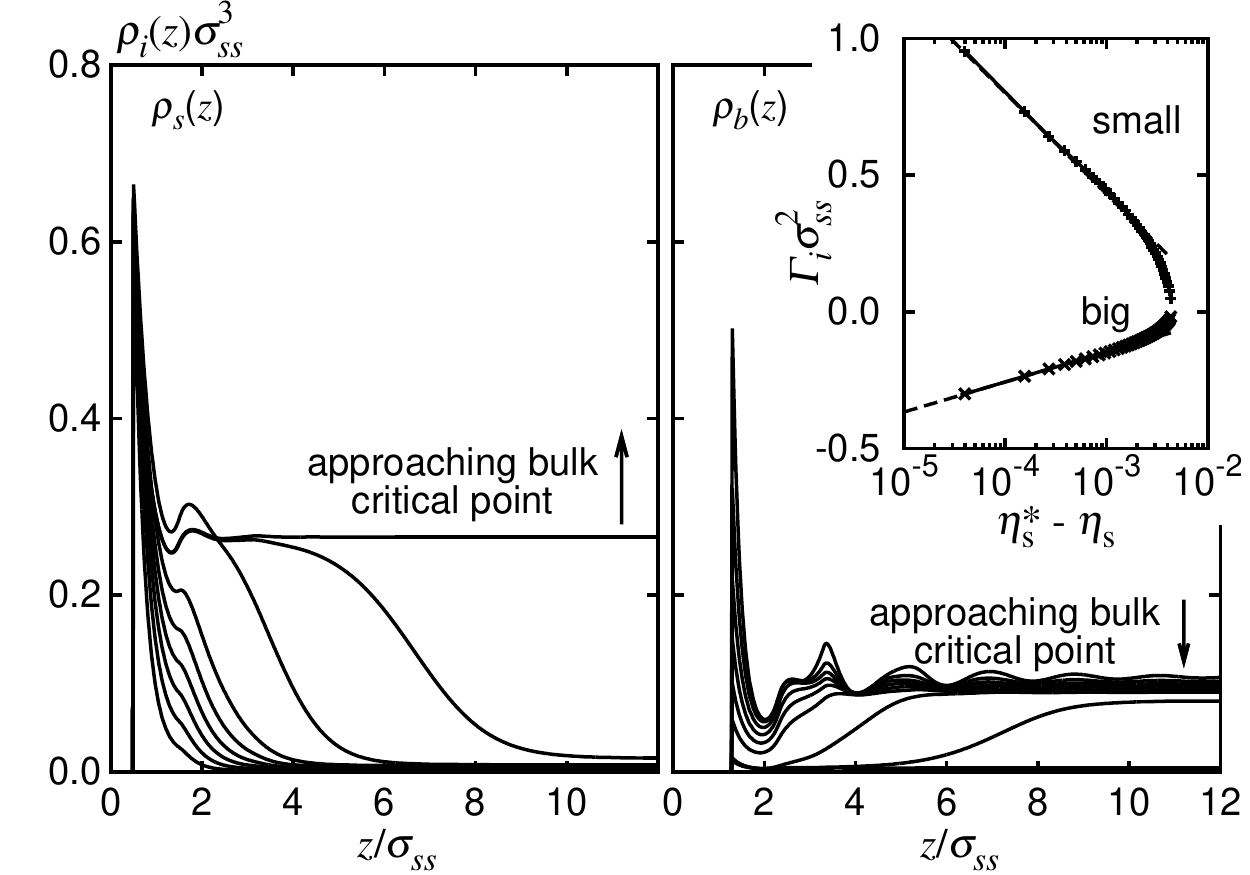}
\caption{\label{fig:1w_profiles_2} Same as
  Fig.~\ref{fig:1w_profiles}(a), but for the $s$-type wall. The
  coexistence curve is traced on the $b$-rich side of the phase
  diagram. As the wetting critical point is approached the smaller
  particles strongly adsorb at the wall, replacing the bigger
  particles and growing a thick film. Below the wetting critical point
  the film is infinitely thick. The inset shows the adsorptions,
  $\Gamma_i$, against the difference in the packing fraction of the
  small species from its value at the wetting critical point,
  $\eta_s^{*}-\eta_s$, on a logarithmic scale, where
  $\eta^*_s=0.0043$.}
\end{figure}

\begin{figure}[t]
\centering
\includegraphics[width=8.0cm]{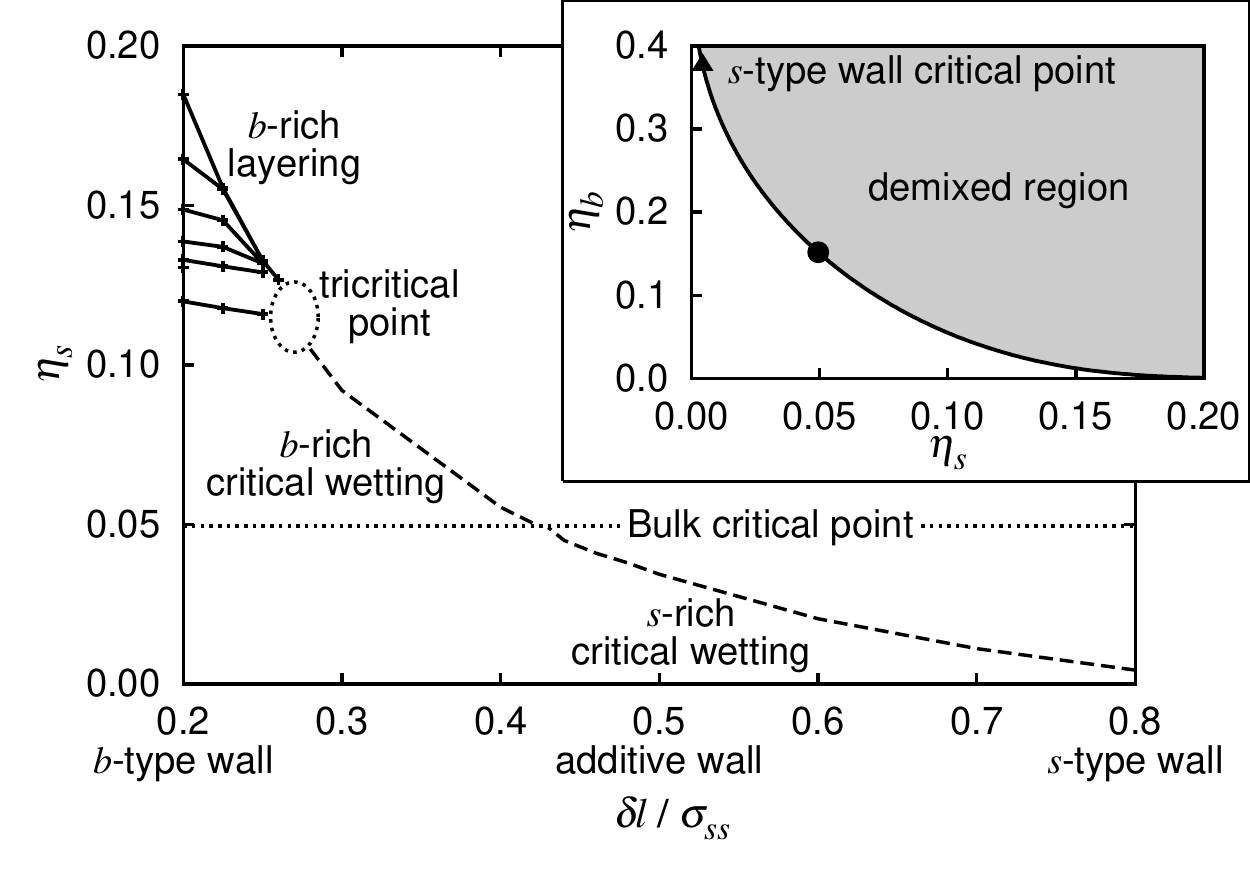}
\caption{\label{fig:pd_deltal_eta1} Value of $\eta_s$ at (i) the
  intercept of the layering transitions with the bulk binodal (solid
  lines) and (ii) the location of the critical wetting transition
  critical point (dashed line), as a function of the scaled wall
  offset $\delta l/\sigma_{ss}$.  As $\delta l/\sigma_{ss}$ is
  increased from 0.2 ($b$-type wall) the layering transitions move
  along the binodal towards the bulk critical point, located at
  $\eta^{\rm crit}_s\simeq0.05$. At $\delta l /\sigma_{ss}\simeq 0.27$
  the layering transitions coalesce and the surface transition becomes
  critical wetting. As $\delta l$ increases towards the additive case,
  the critical wetting transition approaches the bulk critical point.
  Increasing $\delta l$ further, the wetting critical point moves to
  the other side of the binodal. The inset shows the $s$-type wall
  wetting critical point ($\blacktriangle$) in relation to the bulk
  critical point ($\bullet$).}
\end{figure}

For $q=0.5$ and $\Delta=0.2$ the DFT predicts fluid-fluid phase
separation with a critical point at $\eta_s=0.049$, $\eta_b=0.151$ --
see Fig.~2(b).  We start with the $b$-type wall, which we find does
indeed preferentially adsorb the bigger particles. For $b$-rich
statepoints the preferred species is already at the wall and no
surface transitions occur. For $s$-rich statepoints at bulk
coexistence, but far from the bulk critical point, we find that the
small particles dominate the region close to the wall, but that there
is a small amount of adsorption of the bigger particles. To illustrate
this, see the pair of density profiles, $\rho_s(z)$ and $\rho_b(z)$,
furthest from the bulk critical point in
Fig.~\ref{fig:1w_profiles}(a). Reducing $\eta_s$ along the binodal in
the direction towards the bulk critical point, there occurs a series
of discontinuous jumps of the density profiles. The first jump
corresponds to the big particles displacing the small particles from
the wall and forming a layer at a distance $\sigma_{bb}$ away from the
wall, see Fig.~\ref{fig:1w_profiles}(a). Each subsequent jump
corresponds to the adsorption of an extra $b$-rich liquid layer at the
wall. Using the coexistence criteria we have located five distinct
layering transitions. Beyond the fifth transition we find that the
layer rich in the big particles becomes macroscopically thick. We
discuss the possible nature of this transition below. The inset to
Fig.~\ref{fig:1w_profiles}(a) shows the adsorption, $\Gamma_i=\int\dd
z\,\left[\rho_i(z)-\rho_i(\infty)\right]$, of each species $i=s,b$ as
a function of $\eta_s$. Each plateau represents the range of
statepoints along the binodal which have a particular number of
adsorbed layers. The formation of the infinitely thick layer
corresponds to $\Gamma_b$ jumping to $+\infty$, and $\Gamma_s$ to
$-\infty$.  The layering transitions are first-order surface phase
transitions, characterised by a range of coexisting states. In
Fig.~\ref{fig:1w_profiles}(b) we plot the coexistence lines of the
first two transitions in the ($\eta_s,\eta_b$) plane. We find that the
layering transitions intersect the bulk binodal \cite{footnote} and
that they lie very close to the binodal on the $s$-rich side of the
phase diagram. Each transition terminates at a surface critical point,
where the jump in $\Gamma_i$ vanishes. The first layering transition,
where the big particles strongly adsorb at the wall and form the first
layer, is the largest both in terms of the change in the adsorptions
and its size on the phase diagram. Each subsequent transition is
smaller than the previous one.

We next turn to the $s$-type wall. As this preferentially adsorbs the
smaller particles, tracing the bulk coexistence curve on its $b$-rich
side is interesting. For statepoints far from the bulk critical point,
we find that there is some adsorption of the small particles, but that
big particles dominate the region close to the wall, see the pair of
density profiles furthest from the bulk critical point in Fig.~3,
where $\rho_b(z)$ exhibits oscillatory decay that indicates
high-density packing effects. Increasing $\eta_s$ along the binodal in
the direction of the bulk critical point, we find that the small
particles start to adsorb more strongly at the wall, replacing the big
particles. On moving further towards the bulk critical point, a thick
film rich in the small particles grows. No jumps are observed and the
thickness increases continuously (and reversibly) with the state point
up to a wetting critical point, beyond which the film is infinitely
thick, see Fig.~\ref{fig:1w_profiles_2}. Hence we conclude that this
wetting transition is critical. In such a case the adsorption can be
shown \cite{cahn1977critical,dietrich1988phase} to diverge as
$\Gamma_i\propto\log(|\eta^*_s-\eta_s|)$ on the mean-field level,
where $\eta_s^*$ is the value of $\eta_s$ at the wetting critical
point. We find the value of $\eta_s^*$ by fitting $\Gamma_i$ to its
asymptotic form. The inset to Fig.~\ref{fig:1w_profiles_2} compares
the adsorptions to the asymptotic logarithmic form. The location of
the wetting critical point, $\eta_s^*=0.0043$ is shown in relation to
the bulk binodal in the inset to Fig.~\ref{fig:pd_deltal_eta1}.

We next vary the wall offset parameter, $\delta l$, between the two
cases discussed above. Starting with the $b$-type wall, $\delta
l/\sigma_{ss}=0.2$, and increasing $\delta l$ we find that the
location of the layering transitions moves towards the bulk critical
point. In Fig.~\ref{fig:pd_deltal_eta1} we show the value of $\eta_s$
at each of the intersections of a layering transition and the bulk
binodal as a function of $\delta l$. The jump in adsorption at each
layering transition becomes smaller and the extent of the line in the
phase diagram becomes shorter (not shown). Decreasing $\delta l$
further, we find that at $\delta l/\sigma_{ss}\simeq0.27$ the
individual layering transitions bunch up and become indistinguishable
from each other. For smaller $\delta l$ there is a single continuous
wetting transition, where the thickness of the adsorbed $b$-rich layer
grows logarithmically, in a similar manner to the behaviour at the
$s$-type wall described above. We establish the location of the
surface critical point by fitting $\Gamma_i$ to its asymptotic form
and plot the value of $\eta^*_s$ at the wetting critical point in
Fig.~\ref{fig:pd_deltal_eta1}. Increasing $\delta l$ further results
in the location of the wetting critical point moving further along the
bulk binodal towards the bulk critical point so that at $\delta
l/\sigma_{ss}\simeq0.43$ the wetting transition critical point
coincides with the bulk critical point, and the wall is neutral such
that neither species is preferentially adsorbed at the wall.  As
$\delta l$ is increased beyond $0.43$ we find that the wetting
transition moves to the $b$-rich side of the phase diagram. The
additive wall, $\delta l/\sigma_{ss}=0.5$, has a critical wetting
transition, but located very close to the bulk critical point. As
$\delta l$ is increased, the wetting critical point moves further
along the bulk binodal, so that we return back to the $s$-type wall,
$\delta l/\sigma_{ss}=0.8$.

In order to ascertain the generality of our findings, we have
investigated the trends upon changing the model parameters. For size
ratio $q=0.5$ and vanishing wall offset, $\delta l= 0$, we find
layering transitions far from the bulk critical point for a range of
non-additivity parameters $\Delta = 0.1, 0.2, 0.5$.  Adjusting $\delta
l$ towards the case of the additive wall, the layering transitions
move towards the bulk critical point. We also investigated symmetric
mixtures with $q=1$ and $\Delta=0.1$. Clearly, for the additive wall,
$\delta = 0$, there is no preferential adsorption at the wall and
hence no layering transitions. Introducing preferential adsorption via
a non-vanishing wall offset, $\delta l= 0.1, 0.2, 0.3$, layering
transitions occur, and these move away from the bulk critical point
upon increasing~$\delta l$.

In summary, we have shown that the NAHS model exhibits both layering
and critical wetting transitions depending on the hard wall offset
parameter.  We expect this wetting scenario to be general and to occur
in a large variety of systems where steric exclusion is relevant.  A
set of layering transitions had been previously found in the AOV model
at a hard wall \cite{dft_layering}. In these studies the wall
parameter is equivalent to the $b$-type wall. As in these previous
papers the existence of an infinite number of layering transitions is
a possibility within our mean-field DFT treatment. The effects of
fluctuations would be to smear out the higher-order layering
transitions to produce a final `wetting' transition as found here. A
change from a first-order to a critical wetting transition is not
uncommon~\cite{dietrich1988phase}. What is remarkable here is that
tri-critical behaviour can be induced in a purely entropic system by
merely changing a non-additive wall parameter, $\delta l$.  Moreover,
the NAHS model is much less special than the AO model, as here both
species (not only the AO colloids) display short-ranged repulsion and
hence packing effects.  In future work, it would be interesting to see
the effects of non-additivity on wetting in charged systems where
first-order and critical wetting transitions
occur~\cite{oleksyhansen}.

We acknowledge funding by EPSRC under EP/E06519/1 and by DFG under
SFB840/A3.


\begin{thebibliography}{10}

\bibitem{bonn2009wetting}
D. Bonn, J. Eggers, J. Indekeu, J. Meunier, and E. Rolley, Rev. Mod. Phys. {\bf
  81},  739  (2009).

\bibitem{hansen2006tsl}
J.~P. Hansen and I.~R. McDonald, {\em {Theory of Simple Liquids}}, 3rd ed.
  (Academic Press, London, 2006).

\bibitem{roth2010fundamental}
R. Roth, J. Phys.: Condens. Matter {\bf 22},  063102  (2010).

\bibitem{roth2000binary}
R. Roth and S. Dietrich, Phys. Rev. E {\bf 62},  6926  (2000).

\bibitem{oleksyhansen}
A. Oleksy and J. P. Hansen, Mol. Phys. {\bf 104}, 2871 (2006); Mol. Phys. {\bf
  107}, 2609 (2009).

\bibitem{kalcher}
I. Kalcher, D. Horinek, R. Netz, and J. Dzubiella, J. Phys.: Condens. Matter
  {\bf 21}, 424108 (2009); I. Kalcher, J. C. F. Schulz, and J. Dzubiella, Phys.
  Rev. Lett. {\bf 104}, 097802 (2010).

\bibitem{referenceNAHS}
A. Harvey and J. Prausnitz, Fluid Phase Equil. {\bf 48}, 197 (1989); G. Kahl,
  J. Chem. Phys. {\bf 93}, 5105 (1990); L. Woodcock, Ind. Eng. Chem. Re. 2290
  (2010).

\bibitem{asakuraoosawavrij}
S. Asakura and F. Oosawa, J. Chem. Phys. {\bf 22}, 1255 (1954); A. Vrij, Pure
  Appl. Chem. {\bf 48}, 471 (1976).

\bibitem{schmidt2004rfn}
M. Schmidt, J. Phys.: Condens. Matter {\bf 16},  351  (2004).

\bibitem{hopkins2010binary}
P. Hopkins and M. Schmidt, J. Phys.: Condens. Matter {\bf 22},  325108  (2010).

\bibitem{ayadim2010generalization}
A. Ayadim and S. Amokrane, J. Phys.: Condens. Matter {\bf 22},  035103  (2010).

\bibitem{hopkins2010all}
P. Hopkins and M. Schmidt, to be published.

\bibitem{cahn1977critical}
J. Cahn, J. Chem. Phys. {\bf 66},  3667  (1977).

\bibitem{dietrich1988phase}
S. Dietrich, Vol. XII, C. Domb and J.L. Lebowitz, Eds. (Academic Press, London)
   (1988).

\bibitem{dft_layering}
J. M. Brader et. al., J. Phys.: Condens. Matter {\bf 14}, L1 (2002); M.
  Dijkstra and R. van Roij, Phys. Rev. Lett. {\bf 89}, 208303 (2002); P. P. F.
  Wessels, M. Schmidt, and H. L\"owen, J. Phys.: Condens. Matter {\bf 16},
  S4169 (2004).

\bibitem{footnote} For a quantitative comparison of the bulk
  fluid-fluid demixing phase diagram from DFT to simulation results
  for $q=0.1$ and 1, see Fig.~3 of Ref.~\cite{schmidt2004rfn}
  (location of the critical point) and Fig.~4 of
  Ref.~\cite{hopkins2010binary} (binodals).

\end{thebibliography}
\end{document}